\documentclass[conference]{IEEEtran}
\IEEEoverridecommandlockouts

\usepackage{amsfonts}
\usepackage[style=ieee,backend=biber,sorting=none,natbib,url=false,isbn=false,doi=false]{biblatex}
\usepackage{csquotes}
\usepackage{glossaries}
\usepackage{pgfplots}
\usepackage[hidelinks]{hyperref}
\usepackage{subcaption}
\usepackage{tikz}

\pgfplotsset{compat=1.18}
\usetikzlibrary{arrows, shapes}
\tikzset{
    Speaker/.pic={
            \filldraw[fill=white,pic actions]
            (-15pt,0) --
            coordinate[midway] (-front)
            (15pt,0) --
            ++([shift={(-6pt,8pt)}]0pt,0pt) coordinate (aux1) --
            ++(-18pt,0) coordinate (aux2)
            -- cycle
            (aux1) -- ++(0,6pt) -- coordinate[midway] (-back) ++(-18pt,0) -- (aux2);
            \draw (-1mm,-0.4mm) arc (-180:0:1mm);
            \draw (-2mm,-0.6mm) arc (-180:0:2mm);
            \draw (-3mm,-0.8mm) arc (-180:0:3mm);
        }}
\definecolor{mygray}{RGB}{90,90,90}
\definecolor{mycolor3}{RGB}{44,127,184}

\AtEveryBibitem{\clearfield{urlyear}}
\setacronymstyle{long-short}
\newacronym{AFC}{AFC}{acoustic feedback cancellation}
\newacronym{DR}{DR}{dereverberation}
\newacronym[plural={RIRs}]{RIR}{RIR}{room impulse response}
\newacronym{ARMA}{ARMA}{autoregressive moving average}
\newacronym{WOLA}{WOLA}{weighted overlap-add}
\newacronym{STFT}{STFT}{short-time Fourier transform}
\newacronym{ISTFT}{ISTFT}{inverse STFT}
\newacronym{MISO}{MISO}{multiple-input single-output}
\newacronym{SISO}{SISO}{single-input single-output}
\newacronym{GM}{GM}{gain margin}
\newacronym{MSG}{MSG}{maximum stable gain}
\newacronym{LMS}{LMS}{least mean squares}
\newacronym{RLS}{RLS}{recursive least squares}
\newacronym{SDR}{SDR}{signal-to-distortion ratio}
\newacronym{SRR}{SRR}{signal-to-reverberant ratio}
\newacronym{SFR}{SFR}{signal-to-feedback ratio}
\newacronym{SIR}{SIR}{signal-to-interference ratio}
\newacronym{PA}{PA}{public address systems}
\newacronym{HA}{HA}{hearing aids}
\newacronym{WPE}{WPE}{weighted prediction-error}
\newacronym{caf-ctf}{CAF-CTF}{continuous adaptive filter with convolutive transfer function approximation}
\newacronym{CTF}{CTF}{convolutive transfer function approximation}
\newacronym{CAF}{CAF}{continuous adaptive filter}
\newacronym{CD}{CD}{cepstral distance}

\newcommand{\s}{\mathbf{s}}
\renewcommand{\d}{d}
\newcommand{\m}{\mathbf{m}}
\newcommand{\M}{\mathbf{M}}
\renewcommand{\l}{l}
\renewcommand{\L}{\mathbf{L}}
\newcommand{\W}{\mathbf{W}}
\newcommand{\G}{G}
\newcommand{\F}{\mathbf{F}}
\newcommand{\Htf}{\mathbf{H}}

\newcommand{\T}{\mathrm{T}}
\renewcommand{\H}{\mathrm{H}}

\begin{document}

\title{On the Use of Dereverberation for Acoustic Feedback Cancellation \\
    \thanks{This research was carried out at the ESAT Laboratory of KU Leuven, in the frame of Research Council KU Leuven Project C14-21-0075 ”A holistic approach to the design of integrated and distributed digital signal processing algorithms for audio and speech communication devices”, and Aspirant Grant 11PDH24N (for A. Roebben) from the Research Foundation - Flanders (FWO). The scientific responsibility is assumed by its authors.\\
        \textsuperscript{*} These authors contributed equally to this work.}
}

\author{\IEEEauthorblockN{Basil Liekens\textsuperscript{*}, Arnout Roebben\textsuperscript{*}, Toon van Waterschoot and Marc Moonen}
    \IEEEauthorblockA{KU Leuven, Department of Electrical Engineering (ESAT) \\
        STADIUS Center for Dynamical Systems, Signal Processing and Data Analytics \\
        Kasteelpark Arenberg 10, 3001 Leuven, Belgium \\
        \textsuperscript{*}Corresponding authors: \{basil.liekens, arnout.roebben\}@esat.kuleuven.be}
}

\maketitle

\begin{abstract}
    In public address systems and hearing aids, the maximally achievable amplification or gain is limited by acoustic feedback.
    Therefore, in order to be able to apply a higher gain, feedback cancellation methods are required.
    In addition, it is oftentimes also desirable to dereverberate a recorded signal, that is, remove the late reverberation component of the signal, before playing it back.
    In this paper, it is shown that under two mild conditions, the acoustic feedback signal can be written as a reverberant version of the source signal.
    Therefore, it is possible to treat the joint dereverberation and acoustic feedback cancellation problem as a dereverberation-only problem,
    meaning that dereverberation algorithms can be applied to the joint problem.
    Simulations corroborate this finding.
\end{abstract}

\section{Introduction}\label{sec:intro}
In \gls{PA} and \gls{HA}, one of the main tasks is to amplify the signal received at the microphones.
However, one of the factors limiting the maximally achievable amplification or gain is acoustic feedback, i.e. the acoustic coupling between the loudspeaker and microphones may lead to instability when the Nyquist stability criterion is violated \cite{vanwaterschootFiftyYearsAcoustic2011}.
Furthermore, in real-life applications the recorded signal oftentimes also contains reverberation.
Although the early reflections of a \gls{RIR} may improve speech intelligibility, this will not be the case for the late reverberation \cite{naylorSpeechDereverberation2010}.

In the literature several algorithms have been proposed to deal with both phenomena separately.
For \gls{AFC} there exist algorithms that alter the frequency or phase of the microphone signals \cite{vanwaterschootFiftyYearsAcoustic2011}, make use of notch filters to suppress the signal components at unstable frequencies \cite{vanwaterschootFiftyYearsAcoustic2011}, perform system identification to remove the feedback contribution from the microphone signals \cite{sprietAdaptiveFeedbackCancellation2005,romboutsAcousticFeedbackCancellation2006}, or make use of neural networks, either to guide one of the aforementioned methods \cite{zhanDeepPEMAFCImprovedPredictionErrorMethodbased2025} or perform feedback cancellation directly \cite{zhengDeepLearningSolution2022}.
For the \gls{DR} problem, the main idea is to model each microphone signal as a source signal filtered with a \gls{RIR}.
This source-to-microphone \gls{RIR} can then be divided into two contributions: the early reflections and late reverberation.
The goal of \gls{DR} algorithms is therefore to remove the contribution of the late reverberant tail of the \gls{RIR} while retaining the early reflections \cite{delcroixPreciseDereverberationUsing2007,yoshiokaAdaptiveDereverberationSpeech2009,nakataniSpeechDereverberationBased2010}.

However, to the best of our knowledge, no/few algorithms that perform joint \gls{DR} and \gls{AFC} have been proposed in the literature.
Therefore, in this paper, it is shown that under mild conditions, i.e. that the delay in the closed loop is sufficiently long and that the closed loop transfer function can be reasonably approximated by an FIR filter, the feedback component of the microphone signals can be interpreted as a late reverberation component of the source-to-microphone \glspl{RIR}.
This allows to apply \gls{DR} algorithms to perform joint \gls{DR} and \gls{AFC}.
Simulations are presented that demonstrate the effectiveness of the proposed approach.

The rest of this paper is structured as follows: Section \ref{sec:problem_statement} introduces the problem.
Section \ref{sec:afc_dr} shows that under the aforementioned conditions \gls{AFC} can be written as a \gls{DR} problem.
Section \ref{sec:dr} introduces the algorithm used in this paper to perform joint \gls{DR} and \gls{AFC}.
Section \ref{sec:simulations} provides simulation results, with code made available in \cite{liekensIntegratedafcdr2026}.
The paper ends with conclusions in Section \ref{sec:conclusions}.

\section{Problem statement}\label{sec:problem_statement}

\begin{figure}[!b]
    \centering
    \resizebox{.7\columnwidth}{!}{
\begin{tikzpicture}

	\node[draw=none,minimum size=0cm] (t7) {};
	\node[draw=none,minimum size=0cm,below of=t7,node distance=1.0cm] (temp) {};
	\node[draw=black,circle,minimum size=0.7cm,right of=temp,node distance=-0.28cm] (mic1) {};
	\draw ([yshift=10pt]mic1.east) -- ([yshift=-10pt]mic1.east);
	
	\node[draw=none,below of=mic1,node distance=0.8cm] (t1) {};
	\node[draw=none,right of=t1,node distance=0cm] (vd1) {\huge $\vdots$};
	
	\node[draw=black,circle,minimum size=0.7cm,below of=t1,node distance=1cm] (mic2) {};
	\draw ([yshift=10pt]mic2.east) -- ([yshift=-10pt]mic2.east);

    \node[draw=none,below of=mic1,node distance=0.9cm] (t2) {};
    \node[draw=black,left of=t2,minimum width=2cm,node distance=3.7cm,minimum height=2.5cm,align=center,fill=gray!30!white] (Wm) {\Large $\hat{\W}_0(q,k)^\T$};
	\node[draw=none,left of=vd1,node distance=2.3cm] (vd2) {\huge $\vdots$};

    \node[draw=none,below of=mic1,node distance=-1.5cm] (t3) {};
    \node[draw=black,left of=t3,minimum width=1cm,node distance=5.9cm,minimum height=1cm,align=center] (G) {\Large $\G(q,k)$};
    
    \node[draw=none,below of=mic1,node distance=-1.5cm] (t4) {};
    \node[draw=none,left of=t3,minimum width=2cm,node distance=3.2cm,minimum height=1cm,align=center] (Fhat) {};
	\node[draw=none,right of=Fhat,node distance=-0.1cm] (t4b) {};
    \node[draw=none,below of=t4b,node distance=0.7cm,rotate=90] (vd3) {};

    \node[draw=none,right of=mic1,node distance=0.1cm] (t_loudspeaker1) {};
    \pic [scale=0.8,rotate=90,local bounding box=sp1,right of= t_loudspeaker1,node distance=4cm]{Speaker};

    \node[draw=black,right of=Fhat,minimum width=2cm,node distance=4.8cm,minimum height=1cm,align=center] (F) {\Large $\F(q,k)$};
    \node[draw=none,right of=vd3,node distance=4.8cm,rotate=90] (vd4) {\huge $\vdots$};

    \node[draw=gray,below of=Wm,minimum width=3.6cm,node distance=2.85cm,minimum height=2cm,align=center,color=gray] (Wm_update) {\Large Update\\\Large $\hat{\W}(q,k)^\T$};
    \node[draw=none,right of=Wm_update,node distance=1.95cm] (t4c) {};
    \node[draw=none,below of=t4c,node distance=0.1cm,rotate=90,color=gray,rotate=90] (vd5) {\huge $\vdots$};

    \node[draw=none, minimum width=0cm,below of=mic1,node distance=-0.1cm] (t11) {};
	\node[draw=none,right of=t11,node distance=3.6cm] (bm) {};
	\draw [draw=none,xscale=-1] (-28mm,-6mm) arc (90:-90:1mm);
	\draw [draw=none,xscale=-1] (-28mm,-5mm) arc (90:-90:2mm);
	\draw [draw=none,xscale=-1] (-28mm,-4mm) arc (90:-90:3mm);    

    \node[draw=none,above of=mic1,node distance=0.7cm] {\Large $\m[k]$};
    \node[draw=none,above of=mic1,node distance=3.9cm] {\Large $\l[k]$};
    \node[draw=none,right of=bm,node distance=0cm] (p1) {};           

    \node[draw=none,right of=bm,node distance=0.2cm] (p2) {};  
    
    \node[draw=none,right of=Wm,node distance=7.5cm] (Hm_t) {};
	\node[draw=black,above of=Hm_t,minimum width=1.5cm,node distance=0.2cm,minimum height=2.3cm,align=center] (Hm) {\Large $\Htf(q,k)$};

    \node[draw=none,right of=Hm,node distance=1.8cm] (d) {\Large $\d[k]$};

    \draw [->] (Wm) -| (G.south);	
    \draw [->] (G.north) |- (sp1.west);
    \draw [->] (sp1.east) -| (F.north);
    \node [draw=none,minimum width=0cm] (t5) at ([xshift=0.24cm, yshift=0.1cm] mic1.south) {};
    \node [draw=none,minimum width=0cm] (t6) at ([xshift=-0.6cm, yshift=0.13cm] F.south) {};
    \draw [->] (t6) |- (t5);    
    \node [draw=none,minimum width=0cm] (t7) at ([xshift=0.24cm, yshift=0.1cm] mic2.south) {};
    \node [draw=none,minimum width=0cm] (t8) at ([xshift=0.6cm, yshift=0.13cm] F.south) {};
    \draw [->] (t8) |- (t7);
    \node [draw=none,minimum width=0cm] (t9) at ([xshift=0.6cm, yshift=0cm] Fhat.south) {};
    \node [draw=none,minimum width=0cm] (t10) at ([xshift=-0.6cm, yshift=0.cm] Fhat.south) {};
    \draw [->,gray,line width=0.3mm] (Wm_update.north) -- (Wm.south); 
    \node[draw=none,above of=Wm_update,node distance=-1cm] (u1) {};           
    \node[draw=none,below of=Wm_update,node distance=0.1cm] (u2) {}; 
    \node[draw=gray,fill=gray,circle,circle, inner sep=0pt,  minimum size=0.1cm] at ([xshift=-0.95cm]mic1) (dot_gray4) {};
    \node[draw=gray,fill=gray,circle,circle, inner sep=0pt,  minimum size=0.1cm] at ([xshift=-1.2cm]mic2) (dot_gray5) {};
    \node[draw=none,below of=bm,node distance=0cm] (u7) {};           
    \node[draw=none,left of=u7,node distance=1.05cm] (u8) {};           
    \node[draw=none,below of=u8,node distance=0.65cm] (u9) {};     
    \node [draw=none,minimum width=0cm] (u10) at ([xshift=0.24cm, yshift=0.6cm] mic1.south) {};
    \node [draw=none,minimum width=0cm] (u11) at ([xshift=0.9cm, yshift=0.6cm] mic1.south) {};
    \node [draw=none,minimum width=0cm] (u12) at ([xshift=1.1cm, yshift=0.6cm] mic1.south) {};
    \node [draw=none,minimum width=0cm] (u13) at ([xshift=2.1cm, yshift=0.6cm] mic1.south) {};
    \node [draw=none,minimum width=0cm] (u14) at ([xshift=2.3cm, yshift=0.6cm] mic1.south) {};
    \draw [->] (u10.east-|u11.center) -- (u10); 
    \draw [-] (u12.center) -- (u13.center); 
    \draw [-] (u14.center-|Hm.west) -| (u14.center); 
    \node[draw=none,left of=u10,node distance=-0.66cm] (t_upperArc2a) {};           
    \draw[] (t_upperArc2a) arc (180:0:0.1cm);     
    \node[draw=none,left of=u10,node distance=-1.86cm] (t_upperArc3a) {};           
    \draw[] (t_upperArc3a) arc (180:0:0.1cm);    
    \node [draw=none,minimum width=0cm] (u15) at ([xshift=0.24cm, yshift=0.6cm] mic2.south) {};
    \node[draw=none,right of=u15,node distance=1.86cm] (t_upperArc3b) {};       
    \node [draw=none,minimum width=0cm] (u16) at ([xshift=2.23cm, yshift=0.6cm] mic2.south) {};
    \draw [->] (u16) -- (u15.east); 
    \node [draw=none,minimum width=0cm] (u17) at ([xshift=2.29cm, yshift=0.6cm] mic2.south) {};
    \node[draw=none,left of=u7,node distance=1.0cm] (u18) {};           
    \node[draw=none,below of=u18,node distance=-0.2cm] (u19) {}; 
    \draw[] (t_upperArc3b) arc (180:0:0.1cm);     
    \node [draw=none,minimum width=0cm] (k3) at ([xshift=0.24cm, yshift=0.1cm] mic2.south) {};
    \node [draw=none,minimum width=0cm] (k4) at ([xshift=0.24cm, yshift=0.1cm] mic1.south) {};
    \node [draw=none,minimum width=0cm] (k5) at ([xshift=2.1cm, yshift=0.1cm] mic1.south) {};
    \node [draw=none,minimum width=0cm] (k6) at ([xshift=2.3cm, yshift=0.1cm] mic1.south) {};
    \node [draw=none,minimum width=0cm] (k7) at ([xshift=2.5cm, yshift=0.1cm] mic1.south) {};
    \node[draw=none,right of=k4,node distance=1.86cm] (t_upperArc3c) {};       
    \node[draw=none,right of=k4,node distance=2.26cm] (t_upperArc3d) {};       
    \node [draw=none,minimum width=0cm] (k10) at ([xshift=2.7cm, yshift=0.1cm] mic1.south) {};

    \draw [-] (u17.center-|Hm.west) |- (u17.center);

    \draw [->] (mic1.west) -- (mic1-|Wm.east);
    \draw [->] (mic2.west) -- (mic2-|Wm.east);

    \draw [->] (d.west) -- (Hm.east);

    \node[draw=none,above of=Wm_update,node distance=0.6cm] (u1) {};           
    \node[draw=none,below of=Wm_update,node distance=0.6cm] (u2) {}; 
    \draw [->,gray] (dot_gray5) |- (u1-|Wm_update.east); 
    \draw [->,gray] (dot_gray4) |- (u2-|Wm_update.east);     

\end{tikzpicture}
}
    \caption{Illustration of the joint \gls{DR} and \gls{AFC} concept where the filter \(\hat{\W}_0(q,k)\) is designed to perform this integrated task.}
    \label{fig:afc_situation_sketch}
\end{figure}
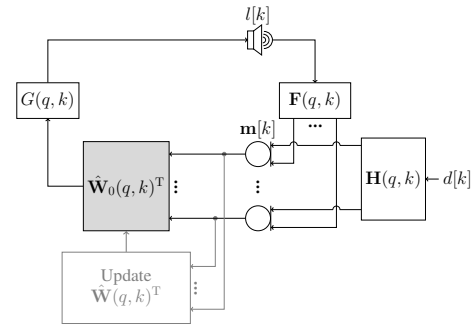

Consider the system depicted in Fig. \ref{fig:afc_situation_sketch} with \(M\) microphones and one loudspeaker.
The corresponding signals are denoted as \(\m[k] \in \mathbb{R}^{M}\) and \(\l[k] \in \mathbb{R}\) with \(k\) denoting the time index.
These multiple microphones can be exploited to improve the performance of the signal processing in the loop.
The loudspeaker signal is obtained from the microphone signals as \(\l[k] = \G(q,k) \hat{\W}_0(q,k)^\T \m[k]\) with \(q^{-1}\) the unit delay operator, i.e. \(q^{-1} d[k] = d[k - 1]\).
Round brackets are used for (matrices of) transfer functions while square brackets denote (matrices of) scalars.
The \gls{MISO} filter \(\hat{\W}_0(q,k)\) will be designed to perform joint \gls{DR} and \gls{AFC}.
The forward path \(\G(q,k)\) then processes the output of this filter by applying the forward path gain and optionally inserting additional delays, i.e. \(\G(q,k) = g q^{-\delta}\), where \(g \in \mathbb{R}\) is the gain and \(\delta \in \mathbb{N}\) is the additional delay \cite{romboutsAcousticFeedbackCancellation2006}.
A single source emits the signal \(\d[k] \in \mathbb{R}\),
which gets convolved with the source-to-microphone \glspl{RIR} \(\Htf(q,k)\) to yield the contribution of the source signal at the microphones \(\s[k] \in \mathbb{R}^M\)
\begin{equation}\label{eq:base_source}
    \s[k] = \left[H_1(q,k) \cdots H_M(q,k)\right]^\T \d[k] = \Htf(q,k) \d[k].
\end{equation}
The \(H_i(q,k)\) are assumed to be FIR filters of order \(L_H\), i.e.
\begin{equation}
    H_i(q,k) = h_i[0,k] + \ldots + h_i[L_H - 1,k] q^{-L_H + 1}.
\end{equation}
These \(H_i(q,k)\) can be split into an early and late component, \(H_{i,e}(q,k)\) and \(H_{i,l}(q,k)\), where
\begin{equation}\label{eq:RIR_early}
    H_{i,e}(q,k) = h_i[0,k] + \ldots + h_i[L_e - 1,k] q^{-L_e + 1},
\end{equation}
\begin{equation}\label{eq:RIR_late}
    H_{i,l}(q,k) = h_i[L_e,k] q^{-L_e} + \ldots + h_i[L_H - 1,k] q^{-L_H + 1},
\end{equation}
with \(L_e\) the number of samples of the \gls{RIR} that constitute the early reflections.
This leads to
\begin{equation}\label{eq:source_contribution_split}
    \begin{split}
        \s[k]  = & \: \s_e[k] + \s_l[k]                                        \\
        =        & \: \left[ H_{1,e}(q,k) \cdots H_{M,e}(q,k) \right]^\T \d[k] \\
        +        & \: \left[ H_{1,l}(q,k) \cdots H_{M,l}(q,k) \right]^\T \d[k] \\
        =        & \: \Htf_e(q,k)\d[k] + \Htf_l(q,k)\d[k].
    \end{split}
\end{equation}
The third contribution in the microphone signal is the feedback signal, defined as \(\F(q,k) \l[k]\) with \(\F(q,k)\) denoting the \glspl{RIR} from the loudspeaker to the microphones, leading to
\begin{equation}\label{eq:base_mic}
    \begin{split}
        \m[k] & = \s[k] + \F(q,k) \l[k] = \s_e[k] + \s_l[k] + \F(q,k)\l[k] \\
              & = \Htf_e(q,k)\d[k] + \Htf_l(q,k)\d[k] + \F(q,k)\l[k].
    \end{split}
\end{equation}
The goal of joint \gls{DR} and \gls{AFC} then consists in designing \(\hat{\W}_0(q,k)\) such that the first component \(\s_e[k]\) is retained while the late reverberation \(\s_l[k]\) as well as the feedback component \(\F(q,k)\l[k]\) are optimally suppressed.

\section{Application of dereverberation to feedback cancellation}\label{sec:afc_dr}
In order to show that \gls{DR} algorithms can be used to perform \gls{AFC} as well,
it is shown that the feedback component in \(\m[k]\) can be considered part of the late reverberation.
The open loop transfer function of the system of Fig. \ref{fig:afc_situation_sketch} is
\begin{equation}\label{eq:base_loudspeaker}
    \l[k] = \frac{\G(q,k)\hat{\W}_0(q,k)^\T}{1 - \G(q,k)\hat{\W}_0(q,k)^\T\F(q,k)}\s[k].
\end{equation}
Combining this with \eqref{eq:base_source} - \eqref{eq:base_mic} then yields
\begin{equation}
    \begin{split}
        m_i[k] = & \, H_i(q,k) \d[k]                                      \\
        +        & \, F_i(q,k) \G(q,k) \hat{\W}_0(q,k)^\T \Htf(q,k) \d[k] \\
        -        & \, H_i(q,k) \G(q,k) \hat{\W}_0(q,k)^\T \F(q,k) \d[k]   \\
        +        & \, \G(q,k) \hat{\W}_0(q,k)^\T \F(q,k) m_i[k],
    \end{split}
\end{equation}
for \(i = 1, \ldots, M\), which is an \gls{ARMA} model for the microphone signal \(m_i[k]\).
This means that every microphone signal \(m_i[k]\) can also be represented as an IIR filtered version of the source signal \(\d[k]\) as
\begin{equation}
    m_i[k] = \sum_{n = 0}^{\infty} C_i[n,k] \d[k-n],
\end{equation}
where \(C_i(q,k)\) is the corresponding IIR filter.
This IIR filter can, akin to \eqref{eq:RIR_early} - \eqref{eq:RIR_late}, be split into
\begin{equation}
    C_{i,e}(q,k) = C_{i,e}[0, k] + \ldots + C_{i,e}[L_e - 1, k]q^{-L_e + 1},
\end{equation}
\begin{equation}
    C_{i,l}(q,k) = C_{i,l}[L_e, k]q^{-L_e} + \ldots,
\end{equation}
where, ideally, \(C_{i,e}(q,k)\) contains the early reflections and \(C_{i,l}(q,k)\) contains both the late reverberation as well as the feedback component.
This will hold true if the combined delay of \(\hat{\W}_0(q,k)\), \(\G(q,k)\) and \(\F(q,k)\) is sufficiently large.
It should be noted that, as this condition concerns the combined delay, this does not necessarily impose any specific conditions solely on the (delay of the) feedback path \(\F(q,k)\) itself.
Indeed, the delay condition can be satisfied by \(\G(q,k)\) which is under control of the system designer.
In practice the sequence of these three operations will oftentimes have a delay of approximately 20 ms \cite{romboutsAcousticFeedbackCancellation2006}
while typically the boundary between the direct path components or early reflections and the late reverberation lies between 8 - 80 ms \cite[Ch. 2]{naylorSpeechDereverberation2010}.
Therefore, if \(L_e\) is chosen sufficiently small, this condition will indeed be satisfied.
It should also be noted that in practice it might be possible that the \gls{MSG} \cite{vanwaterschootFiftyYearsAcoustic2011} can be increased even if not the entire feedback \gls{RIR} is part of the late reverberation.

If in addition the IIR filters \(C_i(q,k)\) can be reasonably approximated by FIR filters, it will be possible to apply \gls{DR} algorithms, in particular inverse filtering-based ones \cite{miyoshiInverseFilteringRoom1988}, to cancel both the late reverberation as well as the feedback component.
The simulation results in Section \ref{sec:simulations} will corroborate this finding.

\section{Dereverberation algorithm}\label{sec:dr}
In the previous section it was shown that, under two mild conditions, it is possible to treat the joint \gls{DR} and \gls{AFC} problem as a \gls{DR}-only problem.
This means that in principle any \gls{DR} algorithm can be used to perform joint \gls{DR} and \gls{AFC}.
In this paper the \gls{WPE} algorithm \cite{yoshiokaAdaptiveDereverberationSpeech2009,nakataniSpeechDereverberationBased2010} will be used.
Since the intended application is \gls{AFC}, a recursive algorithm will be used \cite{yoshiokaAdaptiveDereverberationSpeech2009,drudeNARAWPEPythonPackage2018}.
Due to the fact that the \glspl{RIR} might become long in \gls{PA} applications, time-domain approaches might become prohibitively expensive.
Therefore a \gls{STFT}-domain approach will be used that employs the \gls{CTF} without crossband filters \cite{avargelSystemIdentificationShortTime2007} as is common for the \gls{WPE} algorithm \cite{nakataniSpeechDereverberationBased2010,drudeNARAWPEPythonPackage2018}.

Because of this the time-domain filter \(\hat{\W}_0(q,k)\) will be implemented as a sequence of an analysis filterbank, a frequency-domain filter and a synthesis filterbank.
Let \(n\) and \(\kappa\) denote the frequency bin and frame indices, respectively.
The DFT length used in the \gls{STFT} is denoted by \(N\).

In this method \(K\) previous frames are used to predict the late reverberation in the current frame.
To prevent source signal cancellation, a lag \(\Delta \in \mathbb{N}_{\geq 1}\) is inserted before frames are used for the prediction.
This lag corresponds to a conversion of \(L_e\) samples to \(\Delta\) frames \cite{nakataniSpeechDereverberationBased2010}.
Let \(\M[n, \kappa] \in \mathbb{C}^{M}\) and \(\hat{\W}_\Delta[n, \kappa] \in \mathbb{C}^{MK}\) be the \gls{STFT} representation of \(\m[k]\) and the \gls{WPE} filter in frequency bin \(n\) at time frame \(\kappa\).
Furthermore, define \(\M_{\Delta}[n, \kappa] \in \mathbb{C}^{MK}\) as
\begin{equation}
    \begin{gathered}
        \M_{\Delta}[n, \kappa] = \\
        \left[\M[n, \kappa - \Delta]^\T \cdots \M[n, \kappa - \Delta - K + 1]^\T\right]^\T.
    \end{gathered}
\end{equation}
For the recursive updating of the \gls{WPE} filter an exponentially weighted \gls{RLS} algorithm is used that additionally performs variance normalization \cite{yoshiokaAdaptiveDereverberationSpeech2009,drudeNARAWPEPythonPackage2018}.
Let \(\mathbf{\Phi}[n, \kappa] \in \mathbb{C}^{MK \times MK}\) be the inverse correlation matrix of \(\M_{\Delta}[n, \kappa]\).
Then the update equations for \(\hat{\W}_{\Delta}[n, \kappa]\) are
\begin{equation}\label{eq:wpe_error}
    e[n, \kappa] = M_1[n, \kappa] - \hat{\W}_{\Delta}[n, \kappa]^\H \M_\Delta[n, \kappa],
\end{equation}
\begin{equation}\label{eq:wpe_update_phi}
    \begin{gathered}
        \mathbf{\Phi}[n, \kappa\!+\!1] = \frac{\mathbf{\Phi}[n, \kappa]}{\lambda}                                                                                                                                                           \\
        - \frac{\mathbf{\Phi}[n, \kappa] \M_\Delta[n, \kappa] \M_\Delta[n, \kappa]^\H \mathbf{\Phi}[n, \kappa]}{\sigma_{n, \kappa} \lambda^2 + \lambda \M_\Delta[n, \kappa]^\H \mathbf{\Phi}[n, \kappa] \M_\Delta[n, \kappa]},
    \end{gathered}
\end{equation}
\begin{equation}\label{eq:wpe_update_w}
    \begin{gathered}
        \hat{\W}_\Delta[n, \kappa\!+\!1] = \hat{\W}_\Delta[n, \kappa] \\
        + \frac{\mathbf{\Phi}[n, \kappa\!+\!1]}{\sigma_{n, \kappa}} \M_\Delta[n, \kappa] e[n, \kappa]^*,
    \end{gathered}
\end{equation}
where \(M_1[n, \kappa]\) is the \gls{STFT} representation of the first microphone signal and \(\sigma_{n, \kappa} = \M[n, \kappa]^\H \M[n, \kappa] \slash M\) is used as an estimate of the source power spectral density \cite{drudeNARAWPEPythonPackage2018}.

\section{Simulation results}\label{sec:simulations}
\subsection{Acoustic scenarios}
For the simulations, \gls{RIR}s from the MYRiAD database \cite{dietzenMYRiADMultiarrayRoom2023} were used with a reverberation time of \(0.5\) s.
Speech signals from the \emph{CSTR-VCTK} corpus \cite{yamagishiCSTRVCTKCorpus2019} were used as source signals.
Each trial then consisted of a speech source emitting a \(10\) s signal from one speaker at a sampling frequency of 16 kHz to an array of 4 microphones in the room and another source playing back the loudspeaker signal.
No measurement noise or interfering sources were used in the simulations.
Futhermore, no additional delay in the forward path was added since the \gls{STFT} processing already introduces a delay of \(N\) samples.
The gain in the forward path \(g\) was defined as a \gls{GM} with respect to the minimal \gls{MSG} \cite{vanwaterschootFiftyYearsAcoustic2011} of each of the individual loudspeaker-to-microphone \glspl{RIR} of the system under study.

\subsection{Algorithms}
To asses the \gls{AFC} performance of the \gls{WPE} algorithm, it is compared to a pure \gls{AFC} algorithm.
For reasons that will be elaborated on in Section \ref{subsec:metrics}, it is not possible to directly compare the dereverberation performance of \gls{WPE} in scenarios with and without feedback.

For the pure \gls{AFC} algorithm, a \gls{CAF} is used, which is a widely adopted \gls{AFC} algorithm \cite{sprietAdaptiveFeedbackCancellation2005,guoEvaluationStateoftheArtAcoustic2013}.
It follows a system identification approach using the loudspeaker signal to predict the contribution of the feedback to the microphone signals.
In order to enable a fair comparison with the \gls{WPE} algorithm, the \gls{CAF} is implemented as an \gls{STFT}-domain filter making use of the \gls{CTF} even though it was shown in \cite{dietzenComparativeAnalysisGeneralized2019} that time-domain \gls{DR} algorithms may outperform their \gls{STFT}-domain counterparts.
The resulting algorithm is referred to as the \gls{caf-ctf}.
Furthermore, since the \gls{CAF} acts on each microphone signal separately, the \gls{caf-ctf} is applied to the \gls{SISO} system consisting of the first microphone of the array and the loudspeaker to avoid requiring additional processing in the forward path.

Define \(L[n, \kappa] \in \mathbb{C}\) and \(\hat{\W}_{\mathrm{CAF}}[n, \kappa] \in \mathbb{C}^{L_\mathrm{CAF}}\) as the \gls{STFT} representation of \(\l[k]\) and the corresponding \gls{caf-ctf}.
Also define \(\L_\mathrm{CAF}[n, \kappa] \in \mathbb{C}^{L_\mathrm{CAF}}\) as
\begin{equation}
    \L_\mathrm{CAF}[n, \kappa] = \left[L[n, \kappa] \cdots L[n, \kappa - L_\mathrm{CAF} + 1] \right]^\T.
\end{equation}
Finally, define \(\mathbf{\Psi}[n, \kappa] \in \mathbb{C}^{L_\mathrm{CAF} \times L_\mathrm{CAF}}\) as the inverse autocorrelation matrix of \(\L_\mathrm{CAF}[n, \kappa]\).
Since \gls{WPE} makes use of an \gls{RLS}-type updating \cite[Ch. 10]{haykinAdaptiveFilterTheory2014}, a similar recursive algorithm for \gls{caf-ctf} is used, leading to the following update equations
\begin{equation}\label{eq:caf_ctf_error}
    e_\mathrm{CAF}[n, \kappa] = M_1[n, \kappa] - \hat{\W}_{\mathrm{CAF}}[n, \kappa]^\H \L_{\mathrm{CAF}}[n, \kappa],
\end{equation}
\begin{equation}\label{eq:caf_ctf_update_psi}
    \begin{gathered}
        \mathbf{\Psi}[n,\kappa\!+\!1] = \frac{\mathbf{\Psi}[n, \kappa]}{\lambda}                                                                                                                                                                   \\
        - \frac{\mathbf{\Psi}[n, \kappa] \L_\mathrm{CAF}[n, \kappa] \L_\mathrm{CAF}[n, \kappa]^\H \mathbf{\Psi}[n, \kappa]}{\alpha_{n, \kappa} \lambda^2 + \lambda \L_\mathrm{CAF}[n, \kappa]^\H \mathbf{\Psi}[n, \kappa] \L_\mathrm{CAF}[n, \kappa]},
    \end{gathered}
\end{equation}
\begin{equation}
    \begin{gathered}
        \hat{\W}_\mathrm{CAF}[n,\kappa\!+\!1] = \hat{\W}_\mathrm{CAF}[n,\kappa] \\
        + \frac{\mathbf{\Psi}[n,\kappa\!+\!1]}{\alpha_{n,\kappa}} \L_\mathrm{CAF}[n,\kappa] e_\mathrm{CAF}[n,\kappa]^*.
    \end{gathered}
\end{equation}
The parameter \(\alpha_{n,\kappa}\) is introduced here to discriminate between two cases.
The first one is obtained by setting \(\alpha_{n,\kappa} = 1\) and leads to exponentially weighted \gls{RLS}.
The second sets \(\alpha_{n,\kappa} = |M_1[n,\kappa]|^2\) to additionally perform a variance normalization similar to \gls{WPE},
which is an approach that has already been applied to \gls{AFC} \cite{vanwaterschootDuallyRegularizedRecursive2007,gil-cachoTransformDomainPrediction2012}.

This leads to the following algorithms being compared:
\begin{enumerate}
    \item \gls{caf-ctf} to gauge the \gls{AFC} performance, no normalization, denoted \enquote{CAF-CTF},
    \item \gls{caf-ctf} with normalization, denoted \enquote{nCAF-CTF}.
    \item \gls{WPE} to assess the effectiveness of \gls{DR} algorithms for joint \gls{DR} and \gls{AFC}, denoted \enquote{WPE},
\end{enumerate}
The \gls{STFT} processing made use of a DFT length \(N = 256\) with 50\% overlap, in line with the typical delay in the closed loop of an \gls{AFC} system \cite{romboutsAcousticFeedbackCancellation2006}.
For the analysis and synthesis windows a square-root Hann window was used.
The exponentially weighted \gls{RLS} made use of a forgetting factor \(\lambda = 0.99\).
For \gls{WPE}, \(K\) and \(\Delta\) were set to 7 and 1, respectively.
These parameters were chosen based on preliminary experiments.
In order to compare the two methods with an equal temporal span, \(L_\mathrm{CAF}\) was set to 8.

Since both \gls{WPE} and the (n)\gls{caf-ctf} are making use of \gls{RLS}, the computational complexity for a filter of length \(N\) is identical, namely \(\mathcal{O}(N^2)\) \cite[Ch. 11]{haykinAdaptiveFilterTheory2014}.
However, due to \gls{WPE} being a multichannel method, \(N = M K\) whereas \(N = L_\mathrm{CAF}\) for the (n)\gls{caf-ctf}.

\subsection{Metrics}\label{subsec:metrics}
As mentioned in Section \ref{sec:afc_dr}, the goal of joint \gls{DR} and \gls{AFC} is to only retain the direct path and early reflections while suppressing the contributions of both the late reverberation and feedback signal.
This means that for \gls{WPE} the output after filtering the microphone signals can be written as
\begin{equation}
    \hat{s}_{e,1} = s_{e,1} + s_{l,1} + F_1 \l - \hat{\W}_{0,\mathrm{P}}^\T \m,
\end{equation}
where \(q\) and \(k\) have been omitted for the sake of conciseness.
Using the same notation as in Section \ref{sec:problem_statement}, the filter \(\hat{\W}_{0,\mathrm{P}}(q,k)\) is the time-domain equivalent of the \gls{STFT}-domain prediction filter \(\hat{\W}_\Delta[n, \kappa]\).
As mentioned before, in \gls{WPE} the delay \(\Delta\) is introduced to avoid source signal cancellation meaning the prediction should leave \(s_{e,1}\) intact \cite{nakataniSpeechDereverberationBased2010}.
The residual \(s_{l,1} + F_1 \l - \hat{\W}_{0,\mathrm{P}}^\T \m\) can therefore be considered as interference to the desired signal \(s_{e,1}\), allowing to define a \gls{SIR} metric as
\begin{equation}
    \mathrm{SIR \ [dB]} = 10 \mathrm{log}_{10} \frac{|s_{e,1}|^2}{|s_{l,1} + F_1 l - \hat{\W}_{0,\mathrm{P}}^\T \m|^2},
\end{equation}
where \(q\) and \(k\) have again been omitted for the sake of conciseness.
For the \gls{caf-ctf} a similar reasoning can be applied by replacing \(\hat{\W}_{0,\mathrm{P}}(q,k)^\T \m[k]\) with \(\hat{W}_\mathrm{CAF}(q,k) \l[k]\) where \(\hat{W}_\mathrm{CAF}(q,k)\) is the time-domain equivalent of \(\hat{\W}_\mathrm{CAF}[n, \kappa]\).
It should be noted that due to the correlation between the different signal components in the delayed \(\m[k]\) and the current late reflections it is not possible to split \(\hat{\W}_{0,\mathrm{P}}(q,k)^\T \m[k]\) into constituent parts which would allow to assess the \gls{DR} and \gls{AFC} performance individually.

To quantify the performance improvement, first the output without processing in the loop is computed.
For \gls{WPE} this corresponds to setting \(\hat{\W}_\Delta = \mathbf{0} = \left[0 \cdots 0\right]^\T\).
Similarly, for the \gls{caf-ctf} this corresponds to \(\hat{\W}_{\mathrm{CAF}} = \mathbf{0}\).
Both of these cases are equivalent to the SISO system consisting of the first microphone and the loudspeaker where the same delay as for the STFT processing is introduced in the forward path.

The split between \(H_{i,e}(q,k)\) and \(H_{i,l}(q,k)\), \(L_e\), was set to \(N/2\) samples in correspondence with the choice of \(\Delta = 1\) frame.
This is in line with the boundaries for \(L_e\) given in Section \ref{sec:afc_dr}.

To assess the intelligibility and quality of the signals the \gls{CD} \cite{kinoshitaSummaryREVERBChallenge2016} and eSTOI \cite{jensenAlgorithmPredictingIntelligibility2016} are used, respectively.
The \gls{CD} computes the mean squared distance between the cepstra of short time frames of a clean reference signal and the processed signal, where lower is better.
The eSTOI predicts the intelligibility of the signal based on a clean reference signal and the processed signal and returns a number between zero and one, one being the most intelligible.
The reference signal is the contribution of the early reflections to the microphone signals while the processed signal is obtained through the \gls{ISTFT} of the error signal from \eqref{eq:wpe_error} and \eqref{eq:caf_ctf_error}.

\begin{figure}[!ht]
    \centering
    \begin{subfigure}{0.7\columnwidth}
        \centering
        \includegraphics[width=\textwidth]{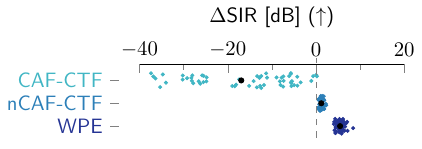}
        \caption{SIR}
        \label{fig:results_SIR}
    \end{subfigure}
    \begin{subfigure}{0.7\columnwidth}
        \centering
        \includegraphics[width=\textwidth]{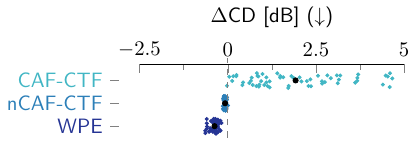}
        \caption{CD}
        \label{fig:results_CD}
    \end{subfigure}
    \begin{subfigure}{0.7\columnwidth}
        \centering
        \includegraphics[width=\textwidth]{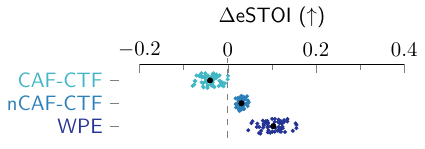}
        \caption{eSTOI}
        \label{fig:results_estoi}
    \end{subfigure}
    \caption{Performance delta when processing in the loop, \gls{WPE} outperforms the \gls{caf-ctf} on all considered metrics. In addition, variance normalization makes a significant difference for the latter.}
    \label{fig:simulation_results}
    \vspace{-12pt}
\end{figure}

\subsection{Discussion}
In Fig. \ref{fig:simulation_results} the three metrics discussed in the previous subsection are shown for a set of scenarios with a \gls{GM} of 6 dB.
Fig. \ref{fig:results_SIR} shows an improvement in \gls{SIR} for \gls{WPE} indicating that, indeed, it manages to perform joint \gls{DR} and \gls{AFC}.
Fig. \ref{fig:results_CD} and Fig. \ref{fig:results_estoi} indicate that this also results in higher quality, more intelligible output as evidenced by a decrease in \gls{CD} and increase in eSTOI.
In addition, all metrics in Fig. \ref{fig:simulation_results} indicate that \gls{WPE} outperforms the (n)\gls{caf-ctf} since the increase in \gls{SIR} and eSTOI and the decrease in \gls{CD} obtained when using \gls{WPE} is larger than that when using the (n)\gls{caf-ctf}.

The poor performance of the regular \gls{caf-ctf} can probably be explained as follows.
First, since the loudspeaker signal and microphone signals are correlated, there will be a residual bias in the filter estimation \cite{sprietAdaptiveFeedbackCancellation2005}.
Second, since \(L_{CAF} = 8\), the temporal span of the filter is 1152 samples which is shorter than the length of the \gls{RIR}, causing undermodeling \cite{romboutsIdentificationUndermodelledRoom2005}.
Furthermore, comparing the n\gls{caf-ctf} to its regular counterpart suggests that the normalization makes a significant difference for both interference suppression as well as the quality and intelligibility of the output.

\begin{figure}[!ht]
    \centering
    \includegraphics[width=.7\columnwidth]{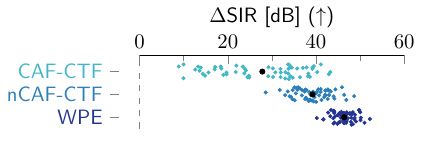}
    \caption{\(\Delta\)\gls{SIR} for -6 dB \gls{GM}, showing that the \gls{caf-ctf} manages to perform \gls{AFC} for unstable systems, but is still outperformed by \gls{WPE}.}
    \label{fig:results_sir_unstable}
    \vspace{-12pt}
\end{figure}

The findings from Fig. \ref{fig:simulation_results} suggest that there is little benefit in using the \gls{caf-ctf} for stable systems.
However, when repeating the experiments of Fig. \ref{fig:simulation_results} with a \gls{GM} of -6 dB Fig. \ref{fig:results_sir_unstable} is obtained, showing that the \gls{caf-ctf} also manages to perform \gls{AFC} in that case.
It is also worth pointing out that the \gls{WPE} still significantly outperforms the (n)\gls{caf-ctf}.

\section{Conclusions}\label{sec:conclusions}
In this paper, it has been shown that under two assumptions acoustic feedback can be considered as a form of reverberation.
First, the feedback loop should contain sufficient delay such that the feedback signal can be considered as part of the late reverberation, a condition that will typically be satisfied.
Second, the IIR filter that can be used to represent the loop transfer function should be reasonably approximated by an FIR filter, a requirement necessary to implement inverse filtering-based \gls{DR} algorithms.
If these two conditions are satisfied, it is possible to apply \gls{DR} algorithms for \gls{AFC}.
Experimental results corroborate this finding:
when comparing \gls{WPE}, an established \gls{DR} algorithm, to the (n)\gls{caf-ctf}, an established \gls{AFC} algorithm, with an equal temporal span, \gls{WPE} offers a better performance in terms of \gls{SIR}, quality measured through \gls{CD} and intelligibility measured through eSTOI.

\printbibliography

\end{document}